\documentclass[preprintnumbers ,amsmath, amssymb, superscriptaddress, longbibliography]{revtex4-1}
\usepackage{amssymb,bm,amsmath}
\usepackage{color, graphicx}

\newcommand{\quotes}[1]{``#1''}
\usepackage{ulem}

\begin{document}

\title{Heaps’ law, statistics of shared components and temporal patterns from a
sample-space-reducing process}

\author{Andrea Mazzolini}
\affiliation{The Abdus Salam International Centre for Theoretical Physics (ICTP), Trieste, Italy}
\author{Alberto Colliva}
\affiliation{Physics Department and INFN, University of Turin, via P. Giuria 1, 10125 Turin, Italy}
\author{Michele Caselle}
\affiliation{Physics Department and INFN, University of Turin, via P. Giuria 1, 10125 Turin, Italy}
\author{Matteo Osella\footnote{To whom correspondence should be addressed.
 Email: mosella@to.infn.it}}
\affiliation{Physics Department and INFN, University of Turin, via P. Giuria 1, 10125 Turin, Italy}

\date{\today}

\begin{abstract}

Zipf's law is a hallmark of several complex systems with a modular structure, such as books composed by words or genomes composed by genes. 
In these component systems, Zipf's law describes the empirical power law distribution of component frequencies.
Stochastic processes based on a sample-space-reducing (SSR) mechanism, in which the number of accessible states reduces as the system evolves, 
have been recently proposed as a simple explanation for the ubiquitous emergence of this law.  
However, many complex component systems are characterized by other statistical patterns beyond Zipf's law, such as a sublinear growth of the 
component vocabulary with the system size, known as Heap's law,  and a specific statistics of shared components. 
This work shows,  with analytical calculations and simulations, that these 
 statistical properties can emerge jointly from a SSR mechanism, thus making it an appropriate parameter-poor representation for component systems. 
Several alternative (and equally simple) models, for example based on the preferential attachment mechanism, can also reproduce Heaps' and Zipf's laws, 
suggesting that additional statistical properties should be taken into account to select the most-likely generative process for a specific system. 
Along this line, we will show that the temporal component distribution predicted by the SSR model is markedly different from the one emerging from the popular
rich-gets-richer mechanism. A comparison with empirical data from natural language indicates that the SSR process can be chosen as 
a better candidate model for text generation based on this statistical property.  
Finally, a limitation of the SSR model in reproducing the empirical ``burstiness'' of word appearances in texts will be pointed out,   
thus indicating a possible direction for extensions of the basic SSR process. 

\end{abstract}

\maketitle

\section{Introduction}

A large number of complex systems have a modular structure. For example,  genomes can be viewed as an assembly of genes, written texts are composed of words, 
and several man-made systems such as softwares or LEGO toys are built starting from basic components.  
Systems with this  modular structure can be described using the  general framework of ~\textit{component systems} \cite{Mazzolini2018}:  
an ensemble of realizations (e.g., genomes, books, LEGO toys) that are simply defined by the statistics of their elementary components (genes, words, LEGO bricks). 
One of the prominent and ubiquitous feature of these complex component systems is a high level of heterogeneity in the usage of components. 
Typically, the component abundances follow the famous Zipf's law. 
This statistical law was first observed~\cite{zipf1949human} and then extensively studied in the context of quantitative linguistics~\cite{li2002zipf, piantadosi2014zipf, altmann2015statistical}, 
and essentially  refers to the empirical fact that word abundances in a written text scale as a power law of the word rank, i.e.,  the position in the list of words sorted by their abundances. 
Moreover, the exponent is usually close to -1. An analogous behaviour has been observed in a huge variety of other complex systems~\cite{newman2005power, mitzenmacher2004brief}, 
from genome composition~\cite{huynen1998frequency}, to firm sizes~\cite{axtell2001zipf}.


Several possible theoretical explanations have been proposed for the ``universal" emergence of Zipf's law~\cite{mitzenmacher2004brief, newman2005power}. 
Stochastic growth with a \textit{preferential attachment} mechanism, i.e., frequent components  have higher probability to further increase their frequency,  
naturally leads to a power-law distribution of component abundances. This rich-gets-richer mechanism is at the basis of many stochastic models 
introduced to describe different component systems such as the Yule-Simon's model~\cite{yule1925mathematical, simon1955class} 
and its different variants introduced in linguistics~\cite{zanette2005dynamics,gerlach2013stochastic},  
the Chinese Restaurant Process~\cite{pitman1997two},  different models based on the Polya's urn scheme~\cite{polya1930quelques, johnson1977urn, tria2014dynamics} 
or on a duplication-innovation dynamics~\cite{rosanova2017}.

Zipf's law has also been interpreted as a sign of critical behaviour, leveraging on the general correspondence between the emergence of power-law statistics 
and criticality in statistical mechanics~\cite{Mora2011}. Following this analogy, Zipf's law can be identified with the probability distribution 
of the microstates, and a power law is expected if the system is close to a critical point. 
This critical state could also emerge as a dynamical consequence of local interactions and without the need of fine tuning as in the  \textit{self-organized-criticality} 
framework~\cite{bak1987self,Mora2011,usher1995dynamic}. 

Without invoking criticality, the \textit{random-group-formation} model instead  tries to explain the widespread emergence of Zipf's law from an entropy maximization principle in the general process
of partitioning of elements into categories (or balls into boxes)~\cite{baek2011zipf}. 
Zipf's law can also emerge from more complex models based on the idea that components have  specific networks of dependencies and that these relations 
 determine their co-occurence in a realization~\cite{Iacopini2018,Mazzolini2018a}.  

More recently, an interesting and simple alternative route for the emergence of Zipf's law has been  proposed~\cite{corominas2015understanding}.    
The candidate mechanims is based on a Sample-Space-Reducing  (SSR) process in which the number of 
accessible states gets smaller as the process dynamics unfolds, defining a ''history-dependent`` random process.  
In the perspective of component systems, the SSR process translates into a stochastic growth process  
in which the number of possible components that can be added to a realization progressively reduces as  
 the realization grows.  
 The composition of a text of natural language has been used as an illustrative example~\cite{corominas2015understanding,thurner2015understanding}.  
Indeed,  in the writing process the usage of a specific word limits the possible choices for the following word due to semantic and syntactic constraints. Therefore,  
the actual number of accessible components reduces with respect to the full vocabulary as a sentence is progressively formed. 
The SSR process provides a minimal (parameter-poor) description for all systems characterized 
by this reduction of the state space during  evolution and can naturally and robustly generate  Zipf's law~\cite{corominas2015understanding,corominas2016extreme}.

However, Zipf's law is not the only statistical regularity that is ubiquitously found in empirical complex component systems. 
A realistic candidate generative model for these systems (e.g., for natural language) is expected to reproduce all these  statistical patterns jointly.  
Therefore, it is necessary to fully characterize the theoretical predictions of the  SSR mechanism with respect to these other statistical properties of component 
systems and compare them with the known empirical trends.   
A clear theoretical understanding of the model predictions can also make the SSR model an effective simple ''null model``  that can be used to disentangle in empirical datasets general 
statistical effects due to the state-space reduction (the main model assumption) from system-specific features due to functional or architectural constraints.   
The general purpose of this work is precisely to fully characterize the statistical properties beyond Zipf's law emerging from the SSR mechanism. 
In particular, we will focus on the statistical features of empirical systems that are detailed below.


A statistical regularity that is often observed in component systems displaying a Zipf's law is Heaps' law. 
This law describes the sublinear growth of the number of different components (i.e. the observed vocabulary) with the system size (i.e., the total number of components), and 
has been observed in several empirical systems 
from linguistics to genomics~\cite{herdan1964quantitative,altmann2015statistical, heaps1978information, cattuto2007semiotic, zhang2009discovering, CosentinoLagomarsino2009}. 
In models based on equilibrium ensembles, such as the random-group-formation model~\cite{baek2011zipf},  the vocabulary is typically a fixed parameter, thus this scaling cannot be addressed straightforwardly.    
On the other hand, stochastic growth models based on preferential attachment can be easily extended by introducing a rate of arrival of new components conveniently chosen to capture 
the empirical Heaps' law~\cite{zanette2005dynamics, CosentinoLagomarsino2009, gerlach2013stochastic, tria2014dynamics}.
The first question we will address is whether Heaps' law can naturally emerge from the SSR mechanism and what is its analytical form depending on the model parameters (Section \ref{sec:heaps}).  

Moreover, Zipf's and Heaps' law are not in general independent. 
In fact, models that build realizations using a simple random sampling of components with relative abundances given by the Zipf's law  
naturally predict Heaps' law, and a precise relation between the exponents of the two power-law behaviours~\cite{van2005formal, lu2010zipf, eliazar2011growth, font2014log}.
A basic assumption of the random sampling procedure is the complete independence between components, and thus the absence of correlations. 
This assumption is in principle violated by the SSR process that could introduce temporal correlations between components due to the temporal evolution of the state-space.  
We will address the question of the relation between the Zipf's and Heaps law obtained by the SSR model by analytical calculations and simulations,  and test whether
the effect of possible correlations can actually be observed in the Heaps-Zipf relation  (Section \ref{sec:heaps}). 


In addition to Heaps' and Zipf's laws, a relevant statistical property of component systems is the distribution of shared components~\cite{Mazzolini2018}. 
This statistics describes the number of components that are in common to a certain number of realizations,
for example the number of words that are in common to a certain number of books. 
This system property is captured by the distribution of occurrences, defined as the fraction of realizations in which a component is present. 
A rare component (small occurrence) appears in a small fraction of realizations, while a common or core component (high occurrence) is present across essentially the whole ensemble of realizations. 
The distribution of occurrences  is well studied in genomics, 
where  the occurrence distribution of the basic components (genes or protein domains) has a peculiar U shape~\cite{touchon2009organised, koonin2008genomics,pang2013universal}. 
This means that there is large number of core and very rare genes with respect to genes shared by an intermediate number of genomes.   
At the same time, the distribution behaviour for small occurrences is well captured by a power-law decay~\cite{pang2013universal}. 
This pattern has gained large attention in the field because of its robustness and  generality across taxonomic levels, giving rise to questions about the 
evolutionary mechanisms at the basis of its origin~\cite{Haegeman2012, Lobkovsky2013}.
Recently, we have extended the analysis of this statistical property to component systems from linguistics and technological systems~\cite{Mazzolini2018} 
and we showed how a random-sampling model that assumes Zipf's law can capture several features of empirical occurrence distributions.   


Here, we will study the occurrence distribution that can be obtained from an ensemble of realizations built with the SSR process (Section \ref{sec:U}). 
In particular, we will show that the SSR model is a good candidate generative model for 
component systems as it can jointly reproduce  Zipf's law,  Heaps' law, and the statistics of shared components often found in empirical systems.  
Classic models based on preferential attachment, such as Simon's model~\cite{gerlach2013stochastic,zanette2005dynamics} or the Chinese Restaurant process~\cite{CosentinoLagomarsino2009}, 
can reproduce Heaps' and Zipf's law, 
but cannot be used to study statistical properties across different realizations, such as the occurrence distribution.  
In fact, in these models the components in a realization are only characterized by their occupation number (their abundances) and are not labelled in any other way. 
Therefore, there is no natural way to compare the presence or absence of a specific component across independent realizations. 



Moreover, there is another critical point of preferential attachment models, especially  as models of text generation. 
In fact, by construction, the components that are selected at the beginning of a realization are expected to show a higher abundance 
with respect to the one selected at the end~\cite{bern2010,baek2011zipf}. 
This is a natural consequence of preferential attachment, since there is a higher probability of re-using components that are present for longer times. 
However, this bias is typically not observed in empirical texts~\cite{bern2010}, as we will further test with an illustrative example.   
The question is if the SSR model also suffers from an analogous bias or if it can represent a more realistic simple generative model for texts. 
To tackle this question, we will  introduce a measure of asymmetry in the temporal (or positional) component distribution  and use it to 
analyze how components of different frequencies are distributed along a realization in the SSR model  
in comparison with results from a model based on preferential attachment (Yule-Simon's model), and with empirical data (Section~\ref{sec:book}).


Finally, a non-trivial correlation pattern in the word occurrences have been observed in natural language, and  
 have been interpreted as an emergent consequence of the language communication purpose, in which  complex ideas and concepts 
have to be projected into a one dimensional sequence of words~\cite{altmann2009beyond,schenkel1993long, alvarez2006hierarchical, altmann2012origin}.  
This correlation pattern can be quantified by looking at the inter-occurrence distance between words, which is highly non-random in data~\cite{altmann2009beyond, font2014log}. 
We will analyze this quantity for realizations of the  SSR model (Section~\ref{sec:book})  showing that, in this case, the model cannot reproduce the empirical trend. 
This discrepancy suggests a possible direction to extend the basic formulation of the SSR model in order to  fully reproduce the complex correlation properties of natural language.

\section{Methods}

\subsection{Definition of the sample-space-reducing process (SSR)}
\label{sec:SSR_def}

\begin{figure}[ht!]
\centering
\includegraphics[width=0.48\textwidth]{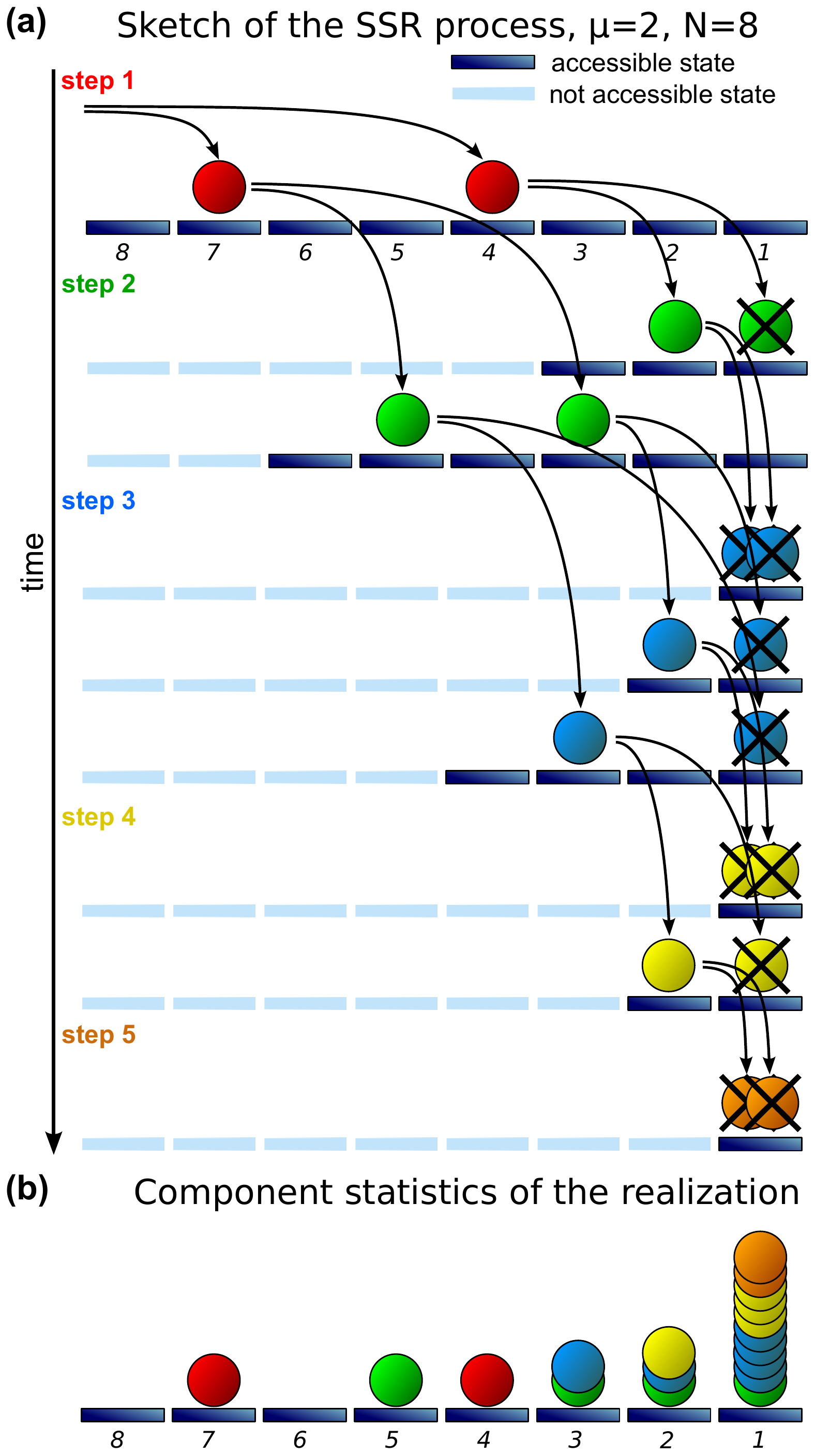}
\caption{
{\bf Schematic representation of the sample-space-reducing (SSR) process.}
a) At the first step, all the $N$ labelled states are accessible, and $\mu$ balls (the $\mu=2$ red balls in the example) 
select among them with uniform probability. At the next step, 
each ball divides into $\mu$ new balls, which jump to states labelled with an index lower than the one of the starting state. 
In the illustrated example, the red ball in state $4$ splits  into two green balls that can only jump to states $\lbrace 3,2,1 \rbrace$.
When a ball  reaches state $1$, it is removed from the process.  Finally, all existing balls will reach this state, thus ending a ``cascade''. 
 The process then restarts with $\mu$ balls thrown over the sample space as in step $1$. 
The SSR process  can be interpreted as a stochastic growth process in which the visited states represent components (e.g., words) that are progressively added to a realization (e.g, a book).  
Therefore, the component statistics of a realization of size $M$ corresponds to the statistics of visited states of the process $\phi_{M}^{\mu}$, as depicted in panel b.
}
\label{fig:sketch}
\end{figure} 
%
%
The basic sample-space-reducing (SSR) process~\cite{corominas2015understanding} is defined as follows.  
A sample space $V$ is composed by $N$  possible states which are labelled and ordered $ \lbrace N, \ldots, 1 \rbrace$. 
A stochastic process is defined over this sample space.  
At the first time step, one of the states is randomly chosen, for instance the state $k$. 
At the following time step, only the last $k-1$ states are accessible, i.e. the sub-set $ \lbrace k-1, \ldots, 1 \rbrace$, and the process selects one of them with uniform probability.  
The procedure is iterated, while the sample space progressively reduces at each iteration, until the ``cascade'' ends with the obligated selection of state $1$. 
After hitting the final state $1$, the process can be re-started  with again $N$  accessible states with equal visiting probability.    
We denote this process as $\phi_{M}$, where $M$ indicates final time step or equivalently the total number of visited states (with their multiplicity). 
During the growth process, the partial time is indicated with $m$, $m \in [1,M]$.
Therefore,  $\phi_{M}$  generates a realization $r$ of size M as a specific ordered sequence of visited states  
$r = (x_1, \ldots, x_M)$ with $x_i \in V$.
To translate this general procedure into a concrete example of a component system, a realization $r$ can be visualized  as a text of natural language. 
The SSR process composes the text by adding at each time step $m$ a word $x_m$ among the possible $N$ word types in the dictionary.   
An ensemble of $R$ realizations can be built as the result of $R$ independent runs of this stochastic process, specifying the final time steps/sizes $\lbrace M_1, \ldots, M_R \rbrace$. 

The basic SSR assumption is that the choice of a word restricts the space of the possible words than can follow it for semantic or structural reasons~\cite{thurner2015understanding}, 
at least for the duration of a cascade.      
The definition of the SSR process  implies  a visiting probability of state $i$ that follows $P_i = i^{-1}$~\cite{corominas2015understanding}. 
This naturally translates for sufficiently long times $M$ (or equivalently for sufficiently large realization sizes)  into an average occupation frequency 
$f_i$ described by the well-known Zipf's law  $f_i \propto i^{-1}$, where $f_i$ corresponds to the normalized number of times that component $i$ has been used in a realization of given size $M$, 
i.e.,  $f_i = \sum_{m=1}^{m=M} \delta_{x_m,i} / M$  for all possible states $i = 1, \ldots, N$.

%
%

The SSR process can be generalized by adding a multiplicative process in order to obtain a visiting probability that follows a power law with arbitrary exponent~\cite{corominas2017sample}. 
This generalization is schematically depicted in Figure~\ref{fig:sketch}.
At the first iteration $\mu$ balls are randomly thrown over a sample space $V$ of $N$ possible states. 
Thus,  $\mu$  states  $\lbrace x_1, \ldots, x_\mu \rbrace$, with $x_k \in \lbrace N, \ldots, 1 \rbrace$, are independently selected with uniform probability among the $N$ possible ones. 
At the next time step,  each of these $\mu$ balls generates again $\mu$ balls that can only fall into states with a lower label, 
following the SSR prescription. 
For example, a ball in state $k$ generates $\mu$ balls that can only bounce on the $k-1$ still accessible states. 
When a ball reaches the final state $1$, it is removed from the process. 
Eventually, all the generated balls will reach this final state,  thus completing the cascade.  
The process can then restart with $\mu$ balls that can randomly choose among the $N$ states.  
We  denote this generalized process as $\phi_{M}^{\mu}$, where $M$ is the number of visited states (or the realization size) and $\mu$ is the free parameter of the multiplicative process. 
In general, for large realizations $M\gg1$,  the number of times the state $i$ is selected by  $\phi_{M}^{\mu}$ 
is simply proportional to $i^{-\mu}$~\cite{corominas2017sample}, thus generalizing the classic Zipf's law. 

The process is not only defined for integer values of $\mu$. In fact, in general, the number of new balls can be extracted from a 
distribution (that has to be defined) with average $\mu$.  
In the following,  we will consider a Poisson distribution with average $\mu$. 
However, we checked numerically that the generalized Zipf's law~\cite{corominas2017sample}, as well as the results we will present for the Heaps' law and for the statistics of 
shared components do not change if  a constant (i.e., no variance) is used for the case of integer $\mu$, or  
if the different prescription presented in ref.~\cite{corominas2017sample} for non-integer $\mu$ values is adopted.

\section{Results}

\subsection{The SSR process naturally generates a sublinear scaling of the number of different components with the realization size (Heaps' law)}
\label{sec:heaps}

Every text of natural language presents a natural ordering of words defined by the reading/writing process from the first $m=1$ word to the end of the text at $m=M$. 
For all component systems whose realizations have this temporal ordering of components, it is possible to evaluate over a single realization 
how the number of different components $h$ grows with the realization size $m$. 
More in general, the same scaling can be analyzed for component systems even without a natural ordering of components (for example for genomes as composed by genes or for LEGO toys)
if the sizes $M$ of the available realizations span a sufficiently large range.  
As discussed in the Introduction,  in several empirical systems this quantity follows a sublinear and approximately power-law function $h(m) \propto m^{\nu}$ (with $\nu < 1$), 
known as Heaps's law~\cite{herdan1964quantitative,altmann2015statistical, heaps1978information, cattuto2007semiotic, zhang2009discovering, CosentinoLagomarsino2009,tria2014dynamics}. 
Each run of the SSR process also generates an ordered sequence of components (or visited states), and the question is what is the predicted scaling of $h(m)$ for this stochastic process.

\begin{figure}[ht!]
\centering
\includegraphics[width=0.8\textwidth]{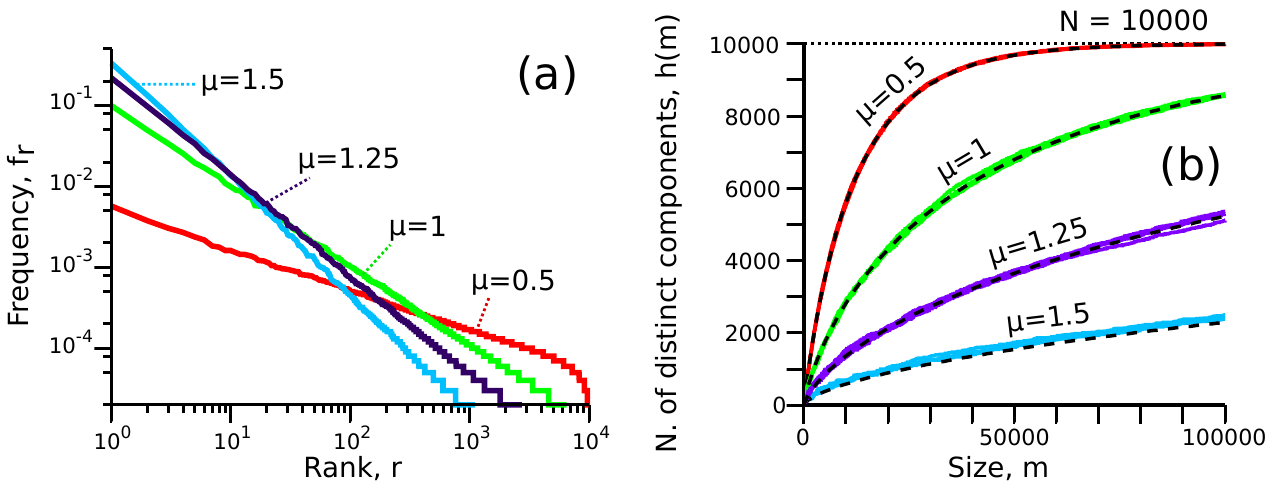}
\caption{{\bf Zipf's law and Heaps' law from the SSR model.}
Panel (a) shows the rank plot of the component frequencies for four realizations of the SSR model with different values of $\mu$. 
Section~\ref{sec:SSR_def} presents in detail the model definition.
The simulations confirm the theoretical expectation in Eq.~\ref{eq:occ_prob}: the power law exponent is simply defined by the value of $\mu$.
Panel (b) shows that the number of different components $h(m)$ grows sublinearly with the realization size.
For each  parameter value, four independent trajectories are reported to give a qualitative idea of the small dispersion around the average.
All trajectories saturate to the asymptotic value $h(m) = N$ (black dotted line), where $N$ is the second free parameter of the model  setting the total number of possible components or  
the vocabulary size. Here, $N=10^4$ for all simulations.
The black dashed lines represent the analytical expression in Eq.~\ref{eq:H(m)}. 
The good overlap with simulations indicates that this analytical approximation can well reproduce the Heaps' law generated by a SSR process.
}
\label{fig:heaps}
\end{figure}

Fig.~\ref{fig:heaps}a reports the rank plots of the component frequencies for different realizations of the SSR process  $\phi_{m}^{\mu}$ with different values of the parameter $\mu$. 
As expected~\cite{corominas2015understanding,corominas2017sample}, they all follow  Zipf's law 

\begin{equation}\label{eq:occ_prob}
p(i) = \frac{i^{-\mu}}{\alpha} \hspace{1cm} \alpha = \sum_{i=1}^{N} i^{-\mu}, 
\end{equation} 

with a power law exponent defined by $\mu$. 

At the same time, Fig.~\ref{fig:heaps}b shows the corresponding scaling of $h(m)$, which increases sublinearly, with a steepness dependent on $\mu$, 
before  saturating at the asymptotic value $h(m) = N$ defined by the total finite number of possible states. 
Therefore,  the behaviour is qualitatively compatible with the empirical Heaps' law suggesting that the SSR process can be a good generative model 
for both  Zipf's and Heaps' laws.

In order to characterize analytically the sublinear growth of $h(m)$ and the precise relation between the two laws in Fig.~\ref{fig:heaps},
we introduce an approximation that neglects the possible correlations that the SSR process can introduce. 
As an example of possible correlations inherent to the SSR process, 
consider  the case of $\phi^{(\mu=1)}_M$, thus with just one ball in the scheme of Fig.~\ref{fig:sketch}.  
The sequence of components selected during a single cascade is strictly decreasing, 
implying that on the time scale of a cascade (which on average last approximately $\log N$~\cite{corominas2015understanding}) 
correlations between sites are present.   
Whether correlations in SSR process are actually sufficiently strong to generate deviations 
from the behaviour of $h(m)$ that can be predicted by neglecting them will be evaluated a posteriori. 

If correlations are neglected, a realization of the SSR process can be 
approximated by assuming that at each time step a component is independently 
drawn with an extraction probability  defined by the visiting probability in Eq.~\ref{eq:occ_prob}. 
This approximation defines a random sampling process with replacement from a fixed number of possible components $N$.

Similar approaches based on random sampling have been previously used to establish the statistical link between Zipf's law and Heaps' law 
 in quantitative linguistics~\cite{van2005formal, lu2010zipf, eliazar2011growth, font2014log}. 
For example, a Poisson process with arrival rates of different components described by Zipf's law
 has been used to compute the sublinear vocabulary growth~\cite{eliazar2011growth}. 
Similarly, Heaps' law has been computed from models based on independent 
 component extractions with \cite{van2005formal} or without \cite{font2014log} replacement, where extraction probabilities were defined by a given (power-law) distribution. 
Our starting assumptions are analogous to the one presented in ref.~\cite{van2005formal}, i.e., a random extraction process with replacement from a power law Zipf's law. 
However,  we will consider the general case of a finite number of possible components $N$, as prescribed by the SSR process we want to approximate,   
while ref.~\cite{van2005formal} focuses on the asymptotic behaviour of Heaps' law in the limit of $N \rightarrow \infty$. 
In general,  previous results will be recovered as specific limiting cases of our framework.

Using the assumption of independent extractions, we can write the probability of choosing for the first time the component $i$ at the step $m$ as
\begin{equation*}
\left(1 - p(i)\right)^{m-1} p(i). 
\end{equation*}
This implies that the component $i$ is selected at least one time  after $m$ steps with probability 
\begin{equation}\label{eq:exist_prob}
q_m(i) = \sum_{l=1}^{m} \left(1-p(i)\right)^{l-1} p(i) = 
1 - \left(1 - p(i)\right)^m . 
\end{equation}
Therefore,  the average value of the number of different components, i.e., the expectation for the  Heaps' law,  can be expressed as 
\begin{equation} \label{eq:H(m)1}
\langle h(m) \rangle = \sum_{i=1}^N q_m(i) = 
N - \sum_{i=1}^N \left(1 - \frac{i^{-\mu}}{\alpha} \right)^m .
\end{equation} 
The above implicit expression  is equivalent to the one presented in \quotes{Lemma 1} of ref.~\cite{van2005formal}. 
In order to get explicit and more intuitive predictions from Eq.~\ref{eq:H(m)1}, some relevant limiting cases can be considered.  
We start by looking at the regime of large realization sizes  $m \gg 1$. 
In this regime, the only relevant terms in the summation are those that satisfy $i^{-\mu}/\alpha \ll 1$. 
Under such a condition, we can take advantage of the logarithm first-order approximation: 
\begin{equation*}
\left(1-\frac{i^{-\mu}}{\alpha}\right)^{m} = 
\exp \left( m \log \left( 1-\frac{i^{-\mu}}{\alpha}\right) \right)
\approx \exp \left( -m \frac{i^{-\mu}}{\alpha} \right) .
\end{equation*}
Substituting all the addends with these exponential forms, and approximating the summation with an integral, one obtains 
\begin{equation*}
\langle h(m) \rangle \approx N - \int_{1}^N \mathrm{d}i \; \exp \left(-m \frac{i^{-\mu}}{\alpha} \right) .
\end{equation*}
This last expression can be evaluated with the change of variables  $z = m i^{-\mu}/\alpha$  
and by making use of the definition of the upper incomplete gamma function 
$\Gamma(n,t) = \int_t^\infty e^{-x} x^{n-1} dx$, thus obtaining the expression:
\begin{equation} \label{eq:H(m)}
\langle h(m) \rangle \approx N - \frac{1}{\mu} \left( \frac{m}{\alpha} \right)^{1/\mu} \Gamma \left( -\frac{1}{\mu}, \frac{m}{N^\mu \alpha} \right) .
\end{equation} 
Note that, even though the special function is defined for a positive first argument, 
it can be extended to negative values by analytic continuation.
Figure \ref{fig:heaps}b shows an extremely  good overlap between the prediction of Eq.~\ref{eq:H(m)} and simulations of the SSR process. 
This good agreement suggests that the Heaps-Zipf relation defined by a random sampling model with no correlations between components is also  satisfied by the SSR model. 
Therefore,  correlations  present in the history-dependent process do not affect significantly the average behavior of the vocabulary growth with the realization size.

A deeper characterization of how the SSR parameters can affect Heaps' law can be obtained by looking at some other limiting cases of Eq.~\ref{eq:H(m)}.
First, we analyze the dynamical approach to  saturation. Given that the total number of possible states $N$ is finite in the SSR process, in the long run ($m \rightarrow \infty$) 
all realization will reach the horizontal asymptote $\langle h(m) \rangle = N$. 
However, the model-parameter values determine the dynamics of this approach to saturation. Indeed, 
Fig. \ref{fig:heaps} (b) shows that  by decreasing $\mu$  the systems discovers new states more quickly, thus making  the Heaps' law steeper.
This observation can be make quantitative by approximating the gamma function in Eq.~\ref{eq:H(m)} 
 with its asymptotic series for $m \gg \alpha N^\mu$. 
 This approximation leads to the expression 
\begin{equation} \label{eq:H_saturation}
\langle h(m) \rangle \approx N \left( 1 - \frac{1}{\mu} \frac{\alpha N^\mu}{m} \exp \left( -\frac{m}{\alpha N^\mu} \right) \right) .
\end{equation}
This exponential form shows that the time scale (or equivalently the size scale) for saturation is defined by 
the quantity 
\begin{equation} \label{eq:crit_size}
\tilde{m} = \alpha N^\mu. 
\end{equation}
When the realization size  is larger than this scale, i.e.,  $m\gg\tilde{m}$,  essentially all the different states have been visited and $\langle h(m) \rangle \approx N$. 
As expected, the time scale of saturation is defined by the total number of states, but the velocity of the exploration of those states depends on the exponent of the Zipf's law.

The opposite regime of $m \ll \tilde{m}$ represents Heaps' law at the beginning of the growth process. 
Note that the Eq.~\ref{eq:H(m)} was derived in the limit of $m \gg 1$, therefore  $N$ have to chosen sufficiently large to satisfy both conditions.
In this case, the gamma function can be reformulated using its recurrence relation  $\Gamma(n+1,t) = n\Gamma(n,t) + t^n e^{-t}$.
Moreover, the  upper gamma can be expressed as the difference between the classical Euler gamma 
and the lower incomplete gamma function, thus  obtaining the following expression 
\begin{equation*} 
\langle h(m) \rangle \approx N \left[ 1 - \exp \left( \frac{m}{\tilde{m}} \right)  
+ \left( \frac{m}{\tilde{m}} \right)^{1/\mu} \left( \Gamma \left( 1-\frac{1}{\mu} \right) - \gamma \left( 1-\frac{1}{\mu}, \frac{m}{\tilde{m}} \right) \right) \right] ,
\end{equation*} 
where $\gamma(n,t) = \int_0^t e^{-x} x^{n-1} dx$.
Applying the limit $\frac{m}{\tilde{m}} \rightarrow 0$ and approximating the exponential function and the lower gamma to the first order, i.e., $\gamma(n,t) \rightarrow t^n / n$, 
we have 

\begin{equation*} 
\langle h(m) \rangle \approx N \left[ \; \frac{m}{\tilde{m}} \; \frac{1}{1-\mu}  + \left( \frac{m}{\tilde{m}} \right)^{1/\mu} \Gamma \left( 1-\frac{1}{\mu} \right) \right] .
\end{equation*} 
This expression indicates that the asymptotic behaviour for $m \ll \tilde{m}$ crucially depends on $\mu$.
When $\mu < 1$, the second term is negligible with respect the first one, while the opposite is true if $\mu > 1$.
The case $\mu = 1$ is singular, but can be evaluated integrating by parts Eq.~\ref{eq:H(m)} and using the definition of 
the exponential integral function, $E_1(z) = \int_z^\infty e^{-x} x^{-1} dx$  and its asymptotic expansion.

We can summarize the results for Heaps' law in the far-from-saturation regime ($\frac{m}{\tilde{m}} \rightarrow 0$) as
\begin{equation} 
\langle h(m) \rangle \approx
\begin{cases}
\left( m (\mu - 1) \right)^{1/\mu} \Gamma \left( 1-\frac{1}{\mu}\right)  & \hspace{0.3cm} \text{for } \mu > 1
\\[5pt]
\frac{m}{\ln{N}} \ln \left( \frac{\ln{N} N}{m} \right) & \hspace{0.3cm} \text{for } \mu = 1 ,
\\[5pt]
m & \hspace{0.3cm} \text{for } \mu < 1 
\end{cases}
\label{eq:h(s)_approx}
\end{equation} 
Here, we used the explicit expressions for the normalization factor $\alpha$, 
which is present in definition of the size scale $\tilde{m}$ (Eq.~\ref{eq:crit_size}). 
Using an integral approximation of the sum in Eq.~\ref{eq:occ_prob}, this factor is 
$\alpha \approx 1/(\mu-1)$ for $\mu > 1$,
$\alpha \approx \ln{N}$ for $\mu = 1$, 
and $\alpha \approx N^{1-\mu}/(1-\mu)$ for $\mu < 1$. 
The expression above fully characterizes the different growth regimes of the number of visited states for the SSR process 
when the realizations are much smaller than the sample space, which is often the case in empirical systems such as texts of natural language. 
The presence of a  transition between a linear growth regime for $\mu < 1$ to a sublinear power-law behavior for $\mu > 1$
is in agreement with a derivation of  Heaps' law  from a Poisson growth process assuming Zipf's law~\cite{eliazar2011growth}.

In conclusion, the SSR model can jointly reproduce Heaps' and Zipf's law, and the link between the two 
laws can be safely calculated by neglecting correlation in the stochastic process.

\subsection{The statistics of shared components from an ensemble of realizations of the SSR process}
\label{sec:U}

\begin{figure}[ht!]
\centering
\includegraphics[width=0.8\textwidth]{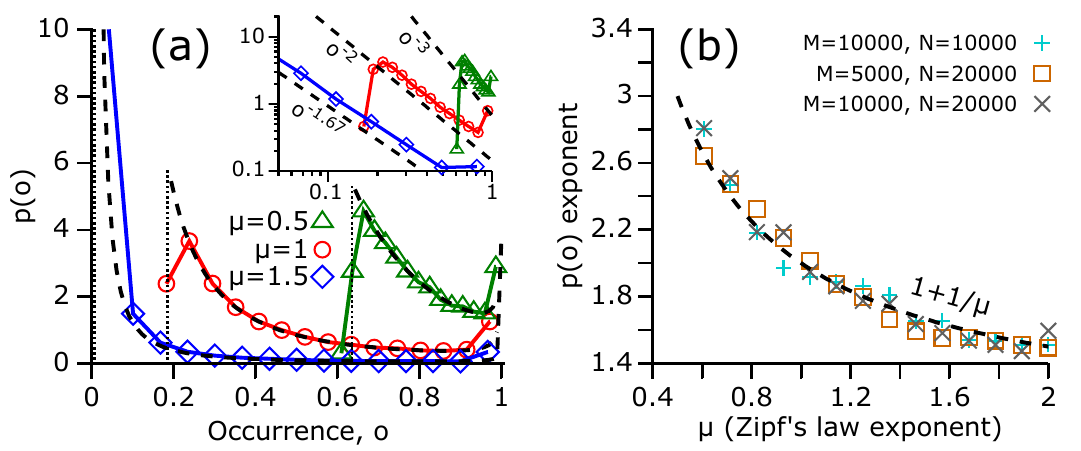}
\caption{{\bf Component occurrence distribution from a SSR process. }
The first panel (a) shows the component occurrence distribution for three ensembles of $R = 1000$ realizations of the SSR process. 
Each ensemble has a different value of $\mu$, while the other two parameters describing the size $M$ of realizations and the number $N$ of possible states are fixed  ($N =M= 10^4$).  
The  distributions obtained by numerical simulations are in good agreement 
with the analytical predictions of Eq.~\ref{eq:plaw_u} (dashed black curves). The distribution left boundaries $o_{left}$ predicted by Eq.~\ref{eq:occ_boundaries} are indicated with vertical dotted lines.  
The inset shows that the same distributions in double-logarithmic scale display a power law decay with an exponent well described by Eq.~\ref{eq:plaw_u_lim}. 
In panel (b) this exponent is estimated for ensembles generated by stochastic simulations of the  SSR model with different parameter values:
$\mu$ values are reported on the x-axis, while $M$ and $N$ values are indicated in the legend. 
Each dot is obtained through a least square fit of the occurrence distribution.
The fitted region is defined as $[o_{left} + \epsilon_1; o_{right} - \epsilon_2]$, where $o_{left}$ and $o_{right}$ are the  boundaries defined by Eqs.~\ref{eq:occ_boundaries}. 
$\epsilon_1$ and $\epsilon_2$ are two positive arbitrary constants chosen to  select the power-law part of the distribution, thus removing the finite-size cut-off for occurrences near $o_{left}$, 
and the increasing part on the right-tail of the distribution that defines the ``core'' components.
The estimated exponents are compared with the analytical expectation (black dashed line) from Eq.~\ref{eq:plaw_u_lim}, which is independent of $M$ and $N$, showing a good agreement.}
\label{fig:u}
\end{figure}

As anticipated in the Introduction, the SSR process provides an ideal framework to 
investigate statistical patterns of components across different realizations. 
In fact,  given a fixed sample space with $N$ states labelled from $N$ to $1$, it is possible to analyze how many states or components are shared by $R$ independent 
realizations of the process $\phi^{\mu}_M$. In other words, the  occurrence distribution $p(o)$  can be computed, where 
the occurrence $o$ of a state (or component)  is defined as  the fraction of realizations in which the state has been selected.  
Three examples (for three $\mu$ values) of occurrence distributions obtained from an ensemble of SSR realizations  are shown in Fig.~\ref{fig:u}(a).
All the three curves display the characteristic U shape often present in empirical data~\cite{Lobkovsky2013,Mazzolini2018,pang2013universal} due to the presence of a 
peak at low occurrences and a second peak at $o=1$ defining the ``core'' components. 
Moreover, the log-log representation in the inset of Fig.~\ref{fig:u}(a) shows 
that for low occurrences the trend is well approximated by a power-law decay.

 Also in this case, it is possible to derive analytical expectations for the statistics of shared components by neglecting possible correlations in the SSR process. 
 This approximation is again equivalent to a random sampling assumption, in which realizations are obtained by  
  independent extractions of components with probabilities defined by the Zipf's law generated by the SSR process (Eq. \ref{eq:occ_prob}).

Here, the observable we are interested in is the component occurrence, which is defined by   
 the probability that component $i$ is present in a realization of size $M$ as described by Eq.~\ref{eq:exist_prob}. 
Therefore, the average fraction $o_i$ of the $R$ realizations in which the component $i$ is present is given by the expression
\begin{equation} \label{eq:occ}
\langle o_i \rangle = \frac{1}{R} \sum_{j=1}^R q_M(i) = 1 - \left(1 - \frac{i^{-\mu}}{\alpha} \right)^M. 
\end{equation}
Note that here we are considering the most simple case in which the probabilities $q_M(i)$ are identical for each realization, 
i.e., all realizations have the same $M$ and $\mu$. 
Therefore,  $q_M(i)$ do not depend on the index $j$ and the summation is trivial.
In general, the scenario in empirical systems can be much more complicated and better described by an ensemble of 
realizations with different sizes $\lbrace M_j \rbrace$,  and coming from slightly different multiplicative processes, i.e., different   $\lbrace \mu_j \rbrace$.

The expression in Eq.~\ref{eq:occ} represents the 
expected occurrence values for a process of random extractions of components from a fixed power law abundance distribution with exponent $\mu$~\cite{Mazzolini2018}. 
Here, we want to test if this formulation can well approximate the results from a SSR process. 
More specifically,  the analytical formula for the component occurrence distribution can be calculated~\cite{Mazzolini2018} as
\begin{equation} \label{eq:plaw_u}
p(o) = \frac{ \left( 1 - o \right)^{1/M - 1}}{\langle h(M R) \rangle \; \mu M  \alpha^{1/\mu} .
\left( 1 - (1 - o)^{1/M} \right)^{1/\mu + 1}} ,
\end{equation}
This distribution is defined in the interval $[o_{left}, o_{right}]$, where:
\begin{equation} \label{eq:occ_boundaries}
o_{left} = \langle o_N \rangle = 1 - \left(1 - \frac{\langle h(M R) \rangle^{-\mu}}{\alpha} \right)^M \hspace{1cm}
o_{right} = \langle o_1 \rangle = 1 - \left(1 - \frac{1}{\alpha} \right)^M .
\end{equation} 
Note that $\langle h(M R) \rangle$ is the expected number of observed different components in the ensemble,
given by Eq.~\ref{eq:H(m)} using the total system size $M R$. 

This analytical expression is compared with simulations of the SSR process in Fig.~\ref{fig:u}(a) showing that also for the statistics of shared 
components the random sampling approximation can well reproduce the model results. 
Also the analytical expressions of the distribution boundaries in Eq.~\ref{eq:occ_boundaries} (vertical dotted lines) are accurate.  
For the sake of simplicity, the simulations were performed close to the  saturation regime, i.e., with a system size $M R$ is much larger than the critical scale $\tilde{m}=N^\mu \alpha$. 
This allows to simplify the expression $\langle h(M R) \rangle \approx N$ in Eq.~\ref{eq:plaw_u}, as discussed in the previous section (see Eq.~\ref{eq:H_saturation}).

The size scale of saturation $\tilde{m}$ defined by Eq.~\ref{eq:crit_size} plays an important role also in determining the global shape of the occurrence distribution.
In fact, the left boundary in Eq.~\ref{eq:occ_boundaries} close to the saturation regime 
(for example for a large number of realizations $R$, such that $M R \gg \tilde{m}$)
can be simply expressed as $o_{left} \approx 1 - \exp \left( - M / \tilde{m} \right)$.
Therefore, the minimal occurrence coincides with zero only when $M \ll \tilde{m}$. 
On the other hand,  $o_{left}$ approaches the maximal occurrence $1$  if $M \gg \tilde{m}$, 
implying that all the components in the ensemble are present in all the realizations.

Another characteristic feature of the occurrence distribution is the power law decay for rare components reported in the inset of Fig.~\ref{fig:u}(a). 
This behaviour can be understood by looking at the limit $o \ll 1$ and $M \gg 1$ in Eq.~\ref{eq:plaw_u}. In fact, in this limit we have 

\begin{equation} \label{eq:plaw_u_lim}
p(o) \approx \frac{M^{1/\mu}}{\alpha^{1/\mu} \; \mu \; \langle h(M R) \rangle} \; o^{-1/\mu - 1} .
\end{equation}
The expression above gives a very simple prediction 
for the exponent of the power law decay, which depends only on $\mu$. This prediction is verified in Fig.~\ref{fig:u}(b), 
showing that the derived simple relation linking
Zipf's law and the statistics of shared components can be safely applied to the SSR process.

\subsection{Temporal dependence of component distributions in the SSR model, in preferential attachment models, and in empirical texts of natural language }
\label{sec:book}

\begin{figure}[ht!]
\centering
\includegraphics[width=0.7\textwidth]{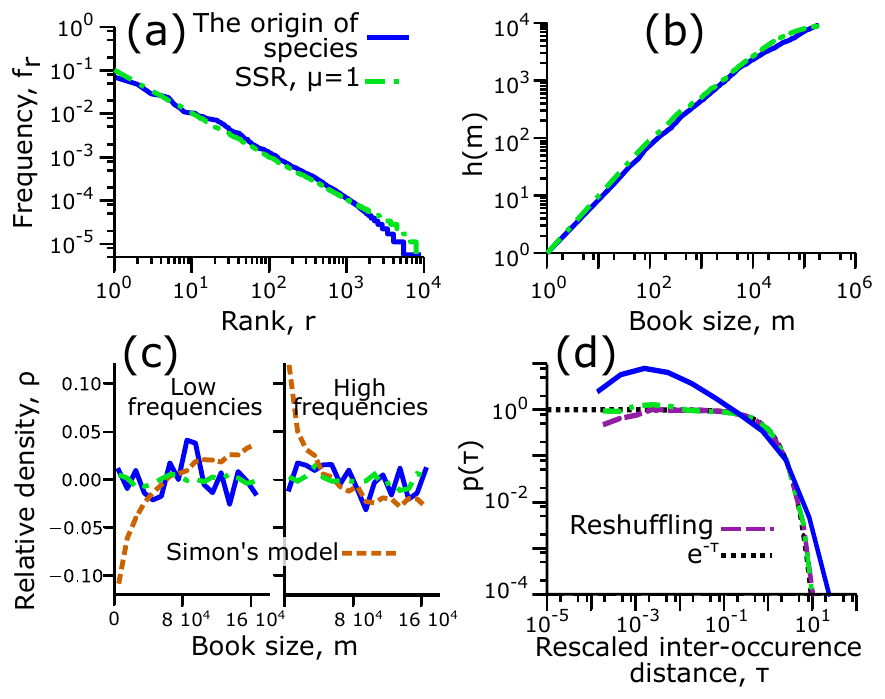}
\caption{{\bf Comparing the SSR model predictions with statistical patterns from Darwin's book  ``On the origin of species'' and from a model based on preferential attachment.} 
 The Zipf's law for the word frequencies in  `` On the origin of species'' (blue line) is compared with the corresponding distribution from a SSR process with $\mu = 1$ (green line-dot curve) in panel (a).   
The number of possible states $N$ and the realization size $M$ are fixed to match the book's vocabulary $N = 9132$ and size $M = 178820$.
With this parameter matching, the SSR model can also well approximate the empirical Heaps' laws as panel (b) shows. 
Panel (c) focuses on the relative density of words (Eq. \ref{eq:density}) $\rho$  for words of low and high frequencies. 
Here, $\rho$ is computed on a moving window of $10^4$ consecutive words. 
The model formulation presented in ref.~\cite{zanette2005dynamics} is used as implementation of a Yule-Simon's process (orange dashed lines), with parameter values $\nu=0.8$ and  $\alpha_0=0.5$ fixed to match the 
empirical Zipf's and Heaps' laws.
For low frequency (in the interval $f \in [10^{-6}, 10^{-4}]$) words, the Simon's model shows a specific increasing relative density along the book (left plot), while the density decreases for high frequency words ($f \in [10^{-3}, 10^{-1}]$).   
Finally, panel (d) displays the inter-occurrence distance distribution $p(\tau)$, with $\tau$ defined by Eq.~\ref{eq:interocc} for $k>1$, 
and evaluated for all words with total abundances between $2$ and $10^3$. 
The  distribution obtained from Darwin's book (blue  line) is compared with the one measured on $20$ realizations of the SSR model 
(green line-dot), and with reshuffled realizations of the model (purple dashed line).  
The theoretical expectation for a random Poisson process is reported as a black dotted line.g.
}
\label{fig:oos}
\end{figure}

We have shown that the SSR model can reproduce Zips's and Heaps' law jointly (Section~\ref{sec:heaps}), making it a good candidate model for many component systems characterized by these laws 
such as texts of natural language. 
The two free model parameters $\mu$ and $N$ can be estimated directly from data since $N$ is the total number of components (e.g., the  text vocabulary), 
while $\mu$ can be set to match the empirical Zipf's law.  
This is shown for the illustrative example of Darwin's  book ``On the origin of species'' in Fig.~\ref{fig:oos}.  
Thee SSR model can produce a realization of the same size $M$ of the text in analysis with a similar word abundance statistics (Fig.~\ref{fig:oos}(a)) 
and,  at the same time,  makes a prediction for the vocabulary growth  that can well approximate the empirical trend (Fig.~\ref{fig:oos}(b)). 
Similar results can be obtained using stochastic growth processes based on a preferential attachment mechanism~\cite{zanette2005dynamics,gerlach2013stochastic,tria2014dynamics,CosentinoLagomarsino2009}
inspired to  the  Yule-Simon's model~\cite{yule1925mathematical, simon1955class}, the Chinese Restaurant Process~\cite{pitman1997two} or the Polya's urn scheme~\cite{polya1930quelques}.  
This makes extremely hard to select the most-likely generative mechanism from these two average behaviours alone.

However, there is a peculiar feature of models based on preferential attachment that emerges by looking at the distribution of components along a realization, thus following its natural temporal order. 
Taking again the example of a text generated with a rich-gets-richer mechanism, rare worlds should be more densely present late in the book, while the opposite should be true for common words. 
In fact,  components that are selected for the first time at the end of the process have a lower chance of being re-selected and trigger the preferential attachment mechanism.   
To make this intuition more quantitative, we introduce a measure of local component density 
and use it to  analyze to what extent this positional (or equivalently temporal) 
bias allows to actually distinguish the SSR model, models based on preferential attachment and empirical data.

Using the notation introduced in Section~\ref{sec:SSR_def}, a realization/book $r$ is an ordered sequence of words/components  $r = (x_1, \ldots, x_M)$, where each 
component instance belongs to the  vocabulary, i.e.,   $x_m \in V = \lbrace 1, \ldots, N \rbrace$.
 A portion of given size of the book $s_{m,\Delta m} = (x_m, \ldots, x_{m + \Delta m})$  can be defined as the sub-sequence of consecutive words from position $m$ to $m + \Delta m$.  
 By definition, we have $s_{ 1, M-1} = r$.  
 We are interested in the density of words of a given frequency class at different positions $m$. 
 Therefore,  we can select the subset of words of the vocabulary  $v_{f_1,f_2} \subset V$  whose frequencies are  in the interval $[f_1, f_2]$. 
Finally, the  relative density $\rho(m, \Delta m, f_1, f_2 )$ of components belonging to the frequency class $v_{f_1,f_2}$ within the portion of the realization $s_{m,\Delta m}$ is 
\begin{equation}\label{eq:density}
\rho(m, \Delta m, f_1, f_2) = \frac{n(s_{m,\Delta m}, \; v_{f_1,f_2})}{\Delta m} - \frac{n(r, \; v_{f_1,f_2})}{M}. 
\end{equation}
 $n(s, v)$ is the number of times that  components belonging to class $v$ appear in $s$, i.e.,  $n(s, v) = \sum_{x \in s} \sum_{c \in v} \delta_{x,c}$. 
The relative density $\rho(m, \Delta m, f_1, f_2)$  measures the difference between the local density in $s_{m,\Delta m}$ and the average density across the whole realization  $n(r, \; v_{f_1,f_2})/M$ 
for components of a given frequency. 
Therefore, this quantity is positive if components of frequency $v$ are enriched at a certain position $m$ of the realization.
The two plots in Fig.\ref{fig:oos}(c) describe how the density of words with low ($f_1 = 10^{-6}$, $f_2 = 10^{-4}$) and 
high ($f_1 = 10^{-3}$, $f_2 = 10^{-1}$) frequency varies  by moving the window $s_{m,\Delta m}$ from the beginning to the end of the book.
The line associated to the Simon's model (as defined in ref.~\cite{zanette2005dynamics}) 
shows  a  clear increasing/decreasing trend for words of  low/high frequencies. This trend quantifies the expected positional (or temporal)  bias inherent to models based on  preferential attachment. 
On the contrary, the SSR model  do not have a specific positional bias for components of different frequencies.  
It simply predicts local fluctuations around the $\rho = 0$ line for all frequency classes. 
A similar trend of local word density is present in the empirical example from  ``On the origin of species'', 
confirming that  words of different frequencies are equally spread across real  texts~\cite{bern2010,baek2011zipf}.

This marked qualitative difference between the SSR process and stochastic processes based on preferential attachment can be used for model selection, 
and suggests that the SSR mechanism is better suited to represent the text generation process with respect to the often-invoked rich-gets-richer scenario.

\subsection{The SSR model cannot reproduce the complex temporal correlation patterns of real texts}

While the average local density of words do not display a specific  temporal trend in empirical texts, it seems to show marked fluctuations (Fig.~\ref{fig:oos}(c)). 
In fact, several studies showed the presence of non-trivial correlation patterns in the temporal distribution 
of words across texts~\cite{altmann2015statistical,altmann2009beyond,altmann2012origin,font2014log}. 
More specifically, the appearance of instances of the same word typically displays a bursty behavior~\cite{altmann2009beyond}, 
which essentially implies the presence of clusters of word instances. 
Intuitively, this behaviour can be traced back to semantic reasons. For example, 
if a character of a novel has a role only in a small part of the story-line, the appearance of his/her name will be localized 
in a corresponding relatively small region of the text.

This statistical pattern can be quantified by looking at the distribution of inter-occurrence distances between words~\cite{altmann2009beyond, font2014log}.  
Given a word $i \in V$ of frequency $f_i$,  
we can compute its $k$th inter-occurrence distance $\tau_i^{(k)}$ as the number of other words 
between its  $(k-1)$th and $k$th appearances normalized by the average 
distance, which is simply given by the inverse frequency $1/f_i$.
In other words, the relative $k$th inter-occurrence distance is defined as 

\begin{equation} \label{eq:interocc}
\tau_i^{(k)} = \left( l_i^{(k)} - l_{i}^{(k-1)} \right) f_i,  
\end{equation}

where $l_i^{(k)} \in \lbrace 1, \ldots, M \rbrace$ represents the position of the $k$th appearance of $i$, with the convention that $l_i^{(0)} = 0$. 
For a completely random distribution of words, the stochastic variable $\tau_i^{(k)}$ follows approximately 
an exponential distribution with average $1$ for any word frequency $f_i$ and  any value of $k$~\cite{font2014log}.  
Therefore, there is a unique null expectation that can be compared to the empirical distribution $p(\tau)$ measured for words of different values of $f_i$ and for different $k$. 
This comparison is reported in Fig.~\ref{fig:oos}(d) for one empirical example, and  confirms that the word inter-occurrence distance  
has a marked excess of short distances with respect to the random expectation (dashed line) for $k>1$. 
 Note that for $k=1$,  $\tau_i^{(1)}$  simply defines the time of first appearance of words, thus it is closely connected to the vocabulary growth (Heaps' law), 
 and the distribution $p(\tau^{(k=1)})$ has been shown to be compatible with the random Poisson expectation~\cite{font2014log}. 
 
The SSR model cannot reproduce the empirical clustering of words, and in fact its prediction for the inter-occurrence distances is well approximated by the exponential random expectation (Fig.~\ref{fig:oos}(d)). 
This means that, at this scale of observation, the ordering of components in SSR realizations is compatible with random ordering. 
In fact, the inter-occurrence distance distribution 
from a SSR realization is not significantly different from its reshuffled version (dashed purple line in Fig.~\ref{fig:oos}(d)), in which the temporal order of components is randomized. 
The  small deviation between the reshuffled realizations and the theoretical exponential expectation at small distances 
is due to the presence of a frequency-dependent lower bound in Eq.~\ref{eq:interocc}, i.e., $\tau_k^{(min)}=f$.
On a finer scale, this equivalence between the SSR model and the random ordering should be violated by the presence of cascades of decreasing order of selected states,  
during which, for example, the same component cannot be selected multiple times (for $\mu\simeq1$).  
However, this effect seems not strong enough to introduce substantial deviations from a random model. 
In conclusion, also for this observable, the correlation structure induced by the SSR model is negligible, and the model predicts that components are approximately 
homogeneously scattered across its realizations.

\section{Discussion}

%
%

The presence of common or universal statistical patterns in complex component systems 
across different fields has attracted a lot of attention~\cite{Mazzolini2018,Holovatch2017,altmann2015statistical,koonin2011}, 
and several alternative mechanisms  have been proposed to be at the origin of these laws.  
Besides the inherent interest in understanding the generative processes at the basis of these emergent patterns, 
simple and parameter-poor models of these systems are also extremely useful as statistical ``null'' models that can be used to disentangle general 
statistical effects from system-specific patterns due to functional or architectural constraints~\cite{Mazzolini2018}. 
This is particularly true in genomics, 
where one is typically interested in identifying the features  
that have been selected by evolution to perform specific biological functions~\cite{koonin2011}. 


In this paper, we have shown that the SSR mechanism can be added to the list of the simple statistical
models that can jointly reproduce Zipf's and Heaps' laws
~\cite{tria2014dynamics, CosentinoLagomarsino2009, gerlach2013stochastic,Iacopini2018,Mazzolini2018a,zanette2005dynamics}. 
Moreover, the SSR model is an appropriate modelling framework to analyze properties of the statistics of shared components, 
which characterize the number of components in common to a given fraction of realizations.  
In particular, the SSR mechanism can naturally produce   
the  U-shaped distribution of occurences that has been observed and intensively studied in genomics~\cite{pang2013universal, Haegeman2012, Collins2012, Lobkovsky2013, Baumdicker2012,Mazzolini2018}. 
This model property marks a relevant difference with respect to commonly used models based on an innovation-duplication dynamics
inspired by the classic Yule-Simon's model~\cite{yule1925mathematical, simon1955class,gerlach2013stochastic,zanette2005dynamics}, 
the Chinese Restaurant Process~\cite{pitman1997two,CosentinoLagomarsino2009} or the Polya's Urn scheme~\cite{polya1930quelques, johnson1977urn, tria2014dynamics}. In fact, 
in these models the components (or states) can be distinguished  only through their occupation numbers, 
while the SSR model, without adding much complexity, has an inherent labelling of the states 
that allows to compare the component composition of independent process realizations.


The precise links between several features of these different statistical patterns generated by the SSR mechanisms 
are well approximated by analytical expressions that neglect possible correlation structures in the model.  
In other words,  a random sampling framework that only assumes the component abundance distribution set by the model seems to capture 
 other statistical model properties, such as the average number of states discovered in time.     
Similarly, the theoretical relation that is often used in linguistics  to connect  Zipf's law and Heaps' law    
is  based on an equivalent random sampling framework~\cite{eliazar2011growth,van2005formal, lu2010zipf,font2014log}. 
Interestingly,  also when these statistical patterns are generated with more complex models 
explicitly based on networks of component dependencies~\cite{Iacopini2018,Mazzolini2018a}, thus with a strong intrinsic correlation structure,
 they do not significantly deviate from the random sampling prediction~\cite{Mazzolini2018a}. 
This surprising phenomenology  suggests that average statistical laws, such as Zipf's and Heaps' laws, do not contain enough information about the microscopic dynamics 
to clearly distinguish between alternative generative mechanisms.  
High-order statistical observables, such as two-point correlations between components~\cite{Mazzolini2018a} or fluctuation scalings~\cite{gerlach2014} 
could thus be necessary to actually select the more appropriate model for a given empirical component system. 

Following this line of reasoning, we introduced a measure of local density of components along a temporally-ordered realization, 
focusing on the specific empirical example of texts of natural language.  
The temporal distribution of components of different frequencies can clearly distinguish realizations of the SSR process with respect to realizations 
built with a preferential attachment mechanism. 
In fact, the rich-gets-richer scenario leads to a high density of low-frequency words at the end of realizations.  
We showed that the SSR model does not introduce this bias,  which is indeed not present in real texts.  
This result identifies the SSR model as a better representation of the text generation process 
with respect to models based on preferential attachment that are often used in this context~\cite{zanette2005dynamics,gerlach2013stochastic}.


While the SSR mechanism seems remarkably effective in reproducing several average empirical trends despite its simplicity, 
it is reasonable to expect that its two-parameter formulation has to be extended to fully capture all statistical properties 
of complex systems such as language. To identify a possible direction for future model extensions, 
we analyzed the inter-occurrence distance distribution in the model and in empirical data.  
In real texts,  this distribution deviates from the uncorrelated scenario of words randomly scattered along the text. 
In fact, it is characterized by an enrichment for short distances  that is due to the tendency of instances of the same word to 
cluster. The presence of topic-dependent structures, epitomized by the subdivision in paragraphs and chapters,
has been suggested as a possible origin of the temporal correlation 
patterns observed in texts~\cite{alvarez2006hierarchical, altmann2012origin}. 
The SSR process clearly does not encode any of these complex features, and consistently we showed that it cannot reproduce the
empirical ``burstiness'' of word appearances.

This limitation of the model suggests a possible direction for future extensions of its basic formulation. 
One possibility would be to include a long-term memory in the state selection process in order to introduce temporal autocorrelations. 
A similar approach has been explored to extend the Yule-Simon's model~\cite{cattuto2006yule}. 
An alternative route could be to consider an underlying spatial organization of the sample space over which the dynamics unfolds. 
Along this line, a model specifically designed for text generation has been studied~\cite{thurner2015understanding}.
The model is inspired to the SSR mechanism, but it is essentially 
 a random walk over empirical networks of words, in which  a link is present if two words 
are found to be consecutive in the text at least one time. While a relation between the network structure 
and the emergence of Zipf's law has been found,  other emergent statistical properties of the model 
and their direct comparison with data are still to be characterized.  
More general models based on the presence of a network of component dependencies have been recently studied~\cite{Mazzolini2018a,Iacopini2018},  
showing that they can  reproduce Heaps' and Zipf's laws. Moreover, an edge-reinforced random walk on a complex dependency network 
can also generate non-random inter-occurrence distance distributions~\cite{Iacopini2018}.   
Identifying the precise relations between these different network-based models,  their key different predictions, the specific role of topology,  
and how these models  are related to the general SSR principle are all interesting directions for future investigations.

\section*{Acknowledgements}

We thank Marco Gherardi, Jacopo Grilli and Marco Cosentino Lagomarsino for useful discussions. 
This work was supported by the "Departments of Excellence 2018 - 2022" Grant 
awarded by the Italian Ministry of Education, University and Research (MIUR) (L. 232/2016)

\bibliography{SSR}

\end{document}